\documentclass[12pt,onecolumn]{IEEEtran}

\usepackage{amssymb, amsmath, graphicx, cite, color, epsfig, subfigure}
\usepackage{xcolor}
\usepackage{caption}
\usepackage{color}
\usepackage{lipsum}
\usepackage{bm} 
\newtheorem{theorem}{Theorem}
\newtheorem{Def1}[theorem]{Definition}

\begin{document}
\title{On the Common Randomness Capacity of a Special Class of Two-way Channels}
\author{\IEEEauthorblockN{Saeed Hajizadeh}} 
\maketitle
\section{Abstratc}
In this paper, we would like to study the common randomness (CR) capacity of intertwined two-way channels, namely those whose marginal channel transition probabilities depends also on the signal they transmit. We bring a few special settings and provide constructive schemes with which the two nodes can agree upon a common randomness. We then provide an outer bound on the CR capacity of intertwined receiver-decomposable (RD) two-way channel and will provide a bound on the cardinality of the available auxiliary variables. We will also show this outer bound is bounded above by  Venkatesan-Anantharam's CR capacity which makes it tight for decomposing two-way setting.

\section{Introduction}
 The concept of CR capacity of different settings has been introduced and studied by Ahlswede and Csiszar \cite{Ahlswede-Csiszar:1998}. The CR capacity of decoupling two-way settings, namely those with general transition function factored as $p(y_1,y_2|x_1,x_2) = p(y_1|x_2)p(y_2|x_1)$,  has been solved by Venkatesan and Anantharam \cite{Venkatesan-Anantharam:1998}. Recently, two-way channel transmission capacities have been studied \cite{Weng-Alajaji:2019}, \cite{Naghizadeh-Berenjian-Razeghi:2015}, \cite{Hajizadeh-Devroye:2014}. We would like to study the CR capacity of intertwined two-way settings, namely those with $p(y_1,y_2|x_1,x_2) = p(y_1|x_1,x_2)p(y_2|x_1,x_2)$. To get a better understanding of what one means by the maximum rate of generating \textit{common} randomness in this setting, and also observe the difference of the CR capacity of a setting with its transmission capacity, we look at an example which at this point it is unknown how complicated finding its CR capacity can be. Consider two terminals, e.g. Alice and Bob, communicate over a special intertwined two-way setting in the absence of any external random sources. In other words, each symbol transmitted through the channel is only a function of the past symbols received. The setting is as follows: Assume Alice's transmitted and received symbols are $X_1$ and $Y_1$ and Bob's transmitted and received symbols are $X_2$ and $Y_2$, respectively. Assuming $0 \leq p_1,q_1,p_2,q_2 \leq \frac{1}{2} $, the transition functions are defined as follows,
\begin{IEEEeqnarray}{rCl}
Y_1|X_1=0,X_2 \sim BSC(p_1) \qquad \qquad Y_1|X_1=1,X_2 \sim BSC(p_2) \\
Y_2|X_1,X_2=0 \sim BSC(q_1) \qquad \qquad Y_2|X_1,X_2=1 \sim BSC(q_2)
\end{IEEEeqnarray}
We now explain different case,
\begin{itemize}
\item[i.] $p_1 = p_2 = q_1 = 0$ and $q_2 = 1/2$: In this case, at the end of each block, Bob can receive some randomness, depending on whether he has sent $X_2 = 1$ or not at the beginning of that block, while Alice receives no randomness. The channel from Bob to Alice is always a $ BSC(0) $ so set $X_1 = 1$ at all blocks. Also to start, set $X_2 = 1$ in the first block and in blocks $b=2,...,n$ Bob sends whatever he receives back to Alice using the-always-$BSC(0)$ backward channel. With this coding scheme it is clear that the first $Y_2=0$ random bit received by Bob stops the random communication from thereon. It can be seen that the average number of random bits per step generated this way is
\begin{IEEEeqnarray}{rCl}
\frac{\sum_{i=1}^{n}i(\frac{1}{2})^i}{n}
\end{IEEEeqnarray}
which vanishes as $n \rightarrow \infty$. However, we can use an even more adaptive coding scheme. In this scheme, Alice always sends $X_1=1$ and Bob starts the first block by sending $X_2 = 1$. If Bob receives $Y_2 = 1$, he sends back $X_2 = 1$ in the next block to agree on that bit with Alice and can still receive a random bit at the end of second block from Alice. This scheme continues until Bob receives $Y_2 = 0$. When he receives $0$ in block $N_1$, he sends it back to Alice in block $N_1+1$ and starts a new block of communication by sending $X_2 =1$. Notice that $N_1$, i.e. the first time Bob receives $0$, is distributed geometrically and $\frac{N_1}{N_1 + 1}$ bits per step is transmitted in the first $N_1 + 1$ blocks. This scheme continues and in the next set of blocks $\frac{N_2}{N_2 + 1}$ random bits per step are generated and agreed upon. Call each $N_i + 1$ blocks of communication a \emph{stage}. Communication takes place in $n$ stages and the number of common random bits per step generated in $n$ stages is 
\begin{IEEEeqnarray}{rCl}
\frac{Z_1 + Z_2 + ... + Z_n}{n}
\end{IEEEeqnarray}
where $Z_i \triangleq \frac{N_i}{N_i + 1}$. Since clearly $Z_i$'s are $i.i.d$, if the number of stages is large, the law of large numbers states that $\mathbb{E}(Z_i)$ common random bits per step are generated and it is easy to see that,
\begin{IEEEeqnarray}{rCl}
\mathbb{E}(Z_i) & = & \mathbb{E}(\frac{N_i}{N_i + 1}) \nonumber \\
& \stackrel{(a)}{=} & \lim_{n \rightarrow \infty}\sum_{j=1}^{n}\frac{j}{j + 1}(\frac{1}{2})^j \nonumber \\
& = & \sum_{j=1}^{\infty}\frac{j}{j + 1}(\frac{1}{2})^j = 0.613706.
\end{IEEEeqnarray}
where $(a)$ follows from the fact that each $N_i$ is distributed geometrically.
Notice that the sum-rate transmission capacity of this setting is at least $1.5$ bits per channel use.
\item[ii.] $p_1 = p_2 = q_2 = 1/2$ and $q_1 = 0$: In this case, set $X_2 = 0$ and as a result, $Y_1$ after each block of communication is a random bit distributed according to $Y_1|X_2=0 \sim BSC(1/2)$ and can send it back noiselessly to terminal 2, Bob. Therefore, after $n$ symbol transmissions, Alice and Bob can agree upon $n - 1$ \textit{uniformly common} bits and thus $R = \frac{\log(2^{n-1})}{n}$ bits per step of communication is achievable. Clearly, $1$ bit per step of communication is the maximum amount of CR that Alice and Bob can agree upon and thus $C=1$ bits/step.
\item[iii.] $p_1 = q_1 = 0$ and $p_2 = q_2 = 1/2$: In this case, Alice and Bob communicate in $n$ blocks. On the first block $X_2$ sends $0$ to set the $Y_2|X_1$ channel as a $BSC(0)$ and $X_1$ sends $1$ to set $Y_1|X_2$ as a $BSC(1/2)$. In the next block, Alice who has receives the bit from $X_2$ randomly retransmitts that bit back to Bob and Bob sends $X_2 = 0$ to open up a noiseless medium to receive the random bit generated in the first block. What $X_2$ sends in the first block is irrelevant. With this scheme, after $n$ transmissions, Alice and Bob can \textit{generate} and \textit{agree upon} $\frac{n}{2}$ bits of randomness which achieves a CR rate of $\frac{\log(2^{\frac{n}{2}})}{n}$ bits per step.

A clear defect in this schemes is Bob's inactiveness in the first block of each transmission cycle. Another approach could be this: In the first block, Alice and Bob each send symbol $1$ and each receive a (possibly different) random bit. To make the \textit{generated} random bit \textit{common}, in the next two blocks, Alice and Bob take turn in sending $0$ to open up the lane for its partner to deliver the random bit noiselessly. With this scheme, $\frac{2n}{3}$ bits are generated and agreed upon in $n$ steps of communication and thus the achievable CR rate is improved to be $2/3$ bits per step. Notice that an outer bound on the CR capacity of this setting is $1$ bit per step. I did not understand how you got 0.75 bit per step. Could you please write it with more detail?
\item[iv.] $p_1 = p_2 = 0$: In this case, $Y_1|X_2 \sim BSC(0)$ at all times thus set, say, $X_1 = 0$. In every block, set $X_{2,i} = Y_{2,i-2}$, $i=1,2,\ldots,n$. To begin the show, set $X_{2,1} = 0$. As a result, $Y_{2,1}=0$ w.p. $q_1$ and $Y_{2,1}=1$ w.p. $1 - q_1$ hence in the second block, the channel switches to $BSC(q_2)$ w.p. $1-q_1$ and stays at $BSC(q_1)$ w.p. $q_1$. Similarly, the channel transition function wavers when it is a $BSC(q_2)$. The whole setting is depicted in Fig. \ref{FS}. So $nC_{FS}$ bits of randomness are generated in this case and can be agreed upon in the other direction. But are they uniformly distributed? Constructively, it is not clear how we can achieve CR in this setting. It turns out that the entropy rate of the output might be the CR of this setting.
\begin{figure}[!ht] 
\includegraphics[width=9cm]{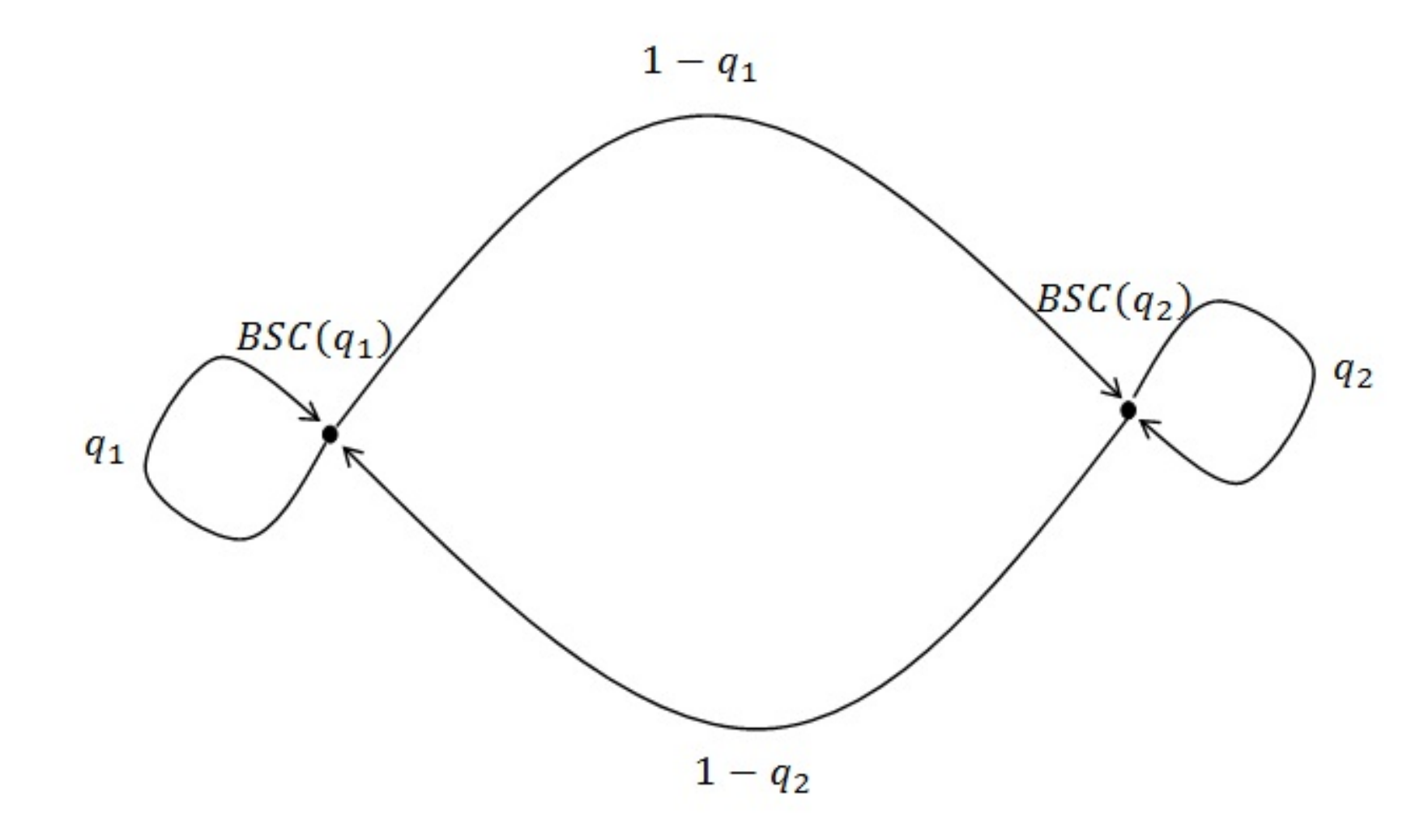}
\centering
\caption{The Finite-State channel modeling the case iv} \label{FS}
\captionsetup{justification=centering}
\end{figure}
\end{itemize}

Now let us define the problem formally.

\section{Definition}
\begin{figure}[!ht] 
\includegraphics[width=7cm]{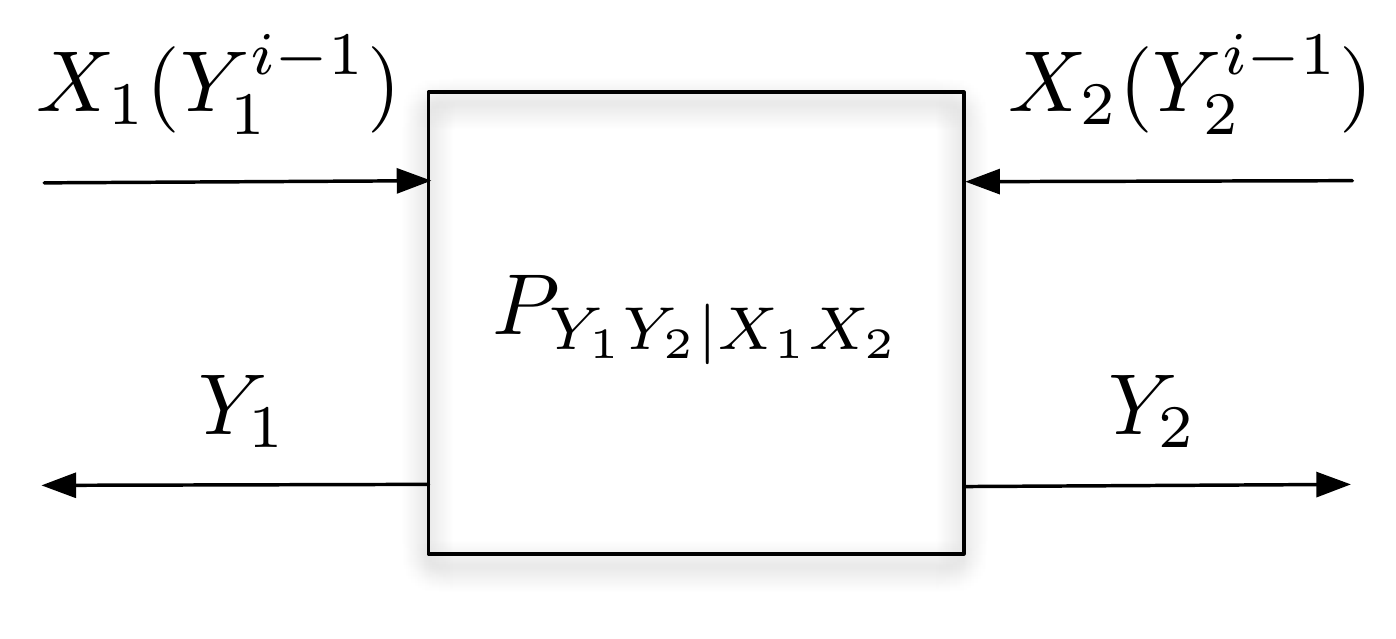}
\centering
\caption{RD two-way channel} \label{RD}
\captionsetup{justification=centering}
\end{figure}

We begin with the definition of a special class of intertwined two-way channels,
\begin{Def1} \label{RDdef}
A special class of two-way channels, depicted as in Fig. \ref{RD}, which we call the receiver-decomposable (RD) two-way channel is described as those in which the channel transition matrix is factored as follows,
\begin{IEEEeqnarray*}{rCl}
p(y_1,y_2|x_1,x_2) = p(y_1|x_1,x_2)p(y_2|x_1,x_2)
\end{IEEEeqnarray*}
\end{Def1}
We desire to find the maximum amount of common randomness the two terminals can agree upon after a block of $n$ uses of the channel. In other words, let the terminals communicate through the channel in $n$ blocks and after gathering the corresponding outputs, each of them can compute a common random output. This random output will take its values from a set, say $\mathcal{K}$. The supremum of the cardinality of $\mathcal{K}$ devided by the number of communication steps, i.e. $\limsup_n \frac{1}{n}K$, is the common randomness capacity of RD two-way channel. To make it more precise, let $X_{1}$, $X_{2}$, $Y_{1}$, and $Y_{2}$ take values from finite sets $\mathcal{X}_1$, $\mathcal{X}_2$, $\mathcal{Y}_1$, and $\mathcal{Y}_2$, respectively. Also let $(f,g)$ be a strategy for transmission by the terminals defined as follows,
\begin{IEEEeqnarray*}{rCl}
f = (f_1,f_2,...,f_n)
\end{IEEEeqnarray*}
and
\begin{IEEEeqnarray*}{rCl}
g = (g_1,g_2,...,g_n)
\end{IEEEeqnarray*}
The terminals would then try to communicate over a certain number of steps, say $n$. In the first step, terminal 1 (T1) sends $f_1 \in \mathcal{X}_1$ and T2 sends $g_1 \in \mathcal{X}_2$. In step $k,2 \leq k \leq n$, T1 constructs from the received symbols in the previous steps, i.e. $Y_1^{k-1}$, the symbol $f_k(Y_1^{k-1}) \in \mathcal{X}_1$. Similarly, T2 sends $g_k(Y_2^{k-1}) \in \mathcal{X}_2$. Come the end of the communication steps, the terminals have the sequence of their received signals, namely $Y_1^n$ and $Y_2^n$. T1 and T2 map $Y_1^n$ and $Y_2^n$ via $\Phi: \mathcal{Y}_1^n \mapsto [1:K] \cup \{e\}$ and $\Psi: \mathcal{Y}_2^n \mapsto [1:K] \cup \{e\}$ for some $\mathcal{K} = \{1,2,...,K\}$ and with $\{e\}$ being the error event in the common randomness generation. We would like to design the strategy such that for every $\lambda > 0$
\begin{IEEEeqnarray}{rCl} \label{UnifCRdef}
\frac{1 - \lambda}{K} \leq \text{Pr}\{ \Phi(Y_1^n) = \Psi(Y_2^n) = l \} \leq \frac{1 + \lambda}{K}~~~\text{for each }l=1,2,...,K.
\end{IEEEeqnarray}
This is called an $(n,K,\lambda)$ strategy to generate common randomness. 
We claim that a non-negative number $R$ is an achievable common randomness if there exists a sequence of $(n,K_n,\lambda_n)$ strategy such that,
\begin{IEEEeqnarray}{rCl}
\liminf_{n \rightarrow \infty} \frac{\log{K_n}}{n} \geq R~~~\text {and}~~~\lim_{n \rightarrow \infty} \lambda_n = 0.
\end{IEEEeqnarray}
We also claim that a non-negative number $R$ is an outer bound on the common randomness capacity if for every strategy $(n,K_n,\lambda)$ we have,
\begin{IEEEeqnarray}{rCl}
\lim_{\lambda \rightarrow 0}\limsup_{n \rightarrow \infty} \frac{\log{K_n}}{n} \leq R.
\end{IEEEeqnarray}
The rate $R$ is said to be an strong outer bound on the common randomness capacity of a setting if for every strategy $(n,K_n,\lambda)$ satisfying (\ref{UnifCRdef}) we have
\begin{IEEEeqnarray}{rCl}
\limsup_{n \rightarrow \infty} \frac{\log{K_n}}{n} \leq R. \qquad \qquad \text{for every}~\lambda \geq 0
\end{IEEEeqnarray}
The strong converse, from the definition, only assures strong convergence of the inner bound and the outer bound but by no means, can necessarily give us a better outer bound.

Notice that in an $(n,K,\lambda)$ protocol, the decoding functions $\Phi$ and $\Psi$ could be taken as the following mappings,
\begin{IEEEeqnarray}{rCl}
\Phi:~\mathcal{X}_1^n \times \mathcal{Y}_1^n \ \mapsto [1:K] \cup \{e\} ~~~~~~~~~\text{and}~~~~~~~~~\Psi:~\mathcal{X}_2^n \times \mathcal{Y}_2^n \ \mapsto [1:K] \cup \{e\}
\end{IEEEeqnarray}
However, it is clear that since each $X_{k,i}$, $k=1,2$, is a function of $Y_k^{i-1}$, mapping from the sequence of received output is sufficient.

Now to find an outer bound on the CR capacity of the RD two-way setting introduced in Definition \ref{RDdef}, let us first bound some entropies. Starting with the joint entropy of the outputs, we have,
\begin{IEEEeqnarray}{rCl}
H(Y_1^n,Y_2^n) & = & \sum_{i=1}^{n}H(Y_{1,i},Y_{2,i}|Y_1^{i-1},Y_2^{i-1}) \nonumber \\
& \stackrel{(a)}{=} & \sum_{i=1}^{n}H(Y_{1,i},Y_{2,i}|Y_1^{i-1},Y_2^{i-1},X_{1,i},X_{2,i}) \nonumber \\
& \stackrel{(b)}{=} & \sum_{i=1}^{n}H(Y_{1,i},Y_{2,i}|X_{1,i},X_{2,i}) \nonumber \\
& \stackrel{(c)}{=} & \sum_{i=1}^{n}H(Y_{1,i}|X_{1,i},X_{2,i}) + H(Y_{2,i}|X_{1,i},X_{2,i}) \nonumber \\
& = & n(H(Y_1|X_1,X_2) + H(Y_2|X_1,X_2))
\end{IEEEeqnarray}
where $(a)$ follows since $X_{k,i}$ is a function of $Y_k^{i-1}$, $k=1,2$, $(b)$ follows from the memorylessness of the setting, and $(c)$ follows from the RD (receiver decompose) property of the channel.
We now bound the marginal entropy of each output,
\begin{IEEEeqnarray}{rCl}
H(Y_1^n) & = & \sum_{i=1}^{n}H(Y_{1,i}|Y_1^{i-1}) \nonumber \\
& \stackrel{(a)}{=} & \sum_{i=1}^{n}H(Y_{1,i}|Y_1^{i-1},X_{1,i}) \nonumber \\
& = & \sum_{i=1}^{n}H(Y_{1,i}|Y_1^{i-1},X_{1,i}) + H(Y_{1,i}|X_{1,i},X_{2,i}) - H(Y_{1,i}|X_{1,i},X_{2,i}) \nonumber \\
& = & \sum_{i=1}^{n}H(Y_{1,i}|X_{1,i},X_{2,i}) + I(X_{2,i};Y_{1,i}|X_{1,i},Y_1^{i-1}) \nonumber \\
& = & \sum_{i=1}^{n}H(Y_{1,i}|X_{1,i},X_{2,i}) + I(X_{2,i};Y_{1,i}|X_{1,i},U_i) \nonumber \\
& = & n(H(Y_1|X_1,X_2) + I(X_2;Y_1|X_1,U))
\end{IEEEeqnarray}
where $(a)$ follows since $X_{k,i}$ is a function of $Y_k^{i-1}$, $k=1,2$ and in the penultimate equality we have defined $U_i \triangleq Y_1^{i-1}$.
Similarly,
\begin{IEEEeqnarray}{rCl}
H(Y_2^n) & = & \sum_{i=1}^{n}H(Y_{2,i}|Y_2^{i-1}) \nonumber \\
& = & \sum_{i=1}^{n}H(Y_{2,i}|Y_2^{i-1},X_{2,i}) \nonumber \\
& = & \sum_{i=1}^{n}H(Y_{2,i}|X_{1,i},X_{2,i}) + I(X_{1,i};Y_{2,i}|X_{2,i},Y_2^{i-1}) \nonumber \\
& = & \sum_{i=1}^{n}H(Y_{2,i}|X_{1,i},X_{2,i}) + I(X_{1,i};Y_{2,i}|X_{2,i},V_i) \nonumber \\
& = & n(H(Y_2|X_1,X_2) + I(X_1;Y_2|X_2,V))
\end{IEEEeqnarray}
where $V_i \triangleq Y_2^{i-1}$.
Notice that this selection of the auxiliary RVs induces the distribution of $p(u,v)p(x_1|u)p(x_2|v)$ and thus the Markov chains $U \rightarrow V \rightarrow X_2$ and $V \rightarrow U \rightarrow X_1$.

We also have,
\begin{IEEEeqnarray}{rCl}
I(Y_1^n;Y_2^n) & = & H(Y_1^n) + H(Y_2^n) - H(Y_1^n,Y_2^n) \nonumber \\
& = & n(I(X_2;Y_1|X_1,U) + I(X_1;Y_2|X_2,V))
\end{IEEEeqnarray}
To bound the cardinality of $U$ and $V$, fix $p(x_1|u)$ and $p(x_2|v)$. Take the following $|\mathcal{X}_1|.|\mathcal{X}_2|+1$ continuous functions of $p(v|u)$,
\begin{displaymath}
g_j(p(v|u)) = \left\{
\begin{array}{lr}
p(x_1,x_2|u) = \sum_{v}p(v|u)p(x_1|u)p(x_2|v) & j = 1,2,\ldots,|\mathcal{X}_1|.|\mathcal{X}_2| - 1 \\
H(Y_1|X_1,U=u) & j = |\mathcal{X}_1|.|\mathcal{X}_2| \\
p(v,x_2,y_2|u) = \sum_{x_1}p(v|u)p(x_1|u)p(x_2|v)p(y_2|x_1,x_2) & j = |\mathcal{X}_1|.|\mathcal{X}_2| + 1
\end{array}
\right.
\end{displaymath} 
The first set of functions preserve $p(x_1,x_2)$ which for the fixed channel transition, preserves $H(Y_1|X_1,X_2)$. The second function preserves $H(Y_1|X_1,U)$ and the last function preserves $p(v,x_2,y_2)$ and thus $H(Y_2|X_2,V)$. Therefore $\mathcal{U}$ suffices to be taken $|\mathcal{U}^{'}| \leq |\mathcal{X}_1|.|\mathcal{X}_2|+1$. Let the corresponding RV of $V$, after replacing $U$ by $U^{'}$ be denoted by $V^{'}$. Now for fixed $U^{'}$ take the following $|\mathcal{X}_1|.|\mathcal{X}_2|$ functions of $p(x_1,x_2|u^{'},v^{'})$,
\begin{displaymath}
h_j(p(x_1,x_2|u^{'},v^{'})) = \left\{
\begin{array}{lr}
p(x_1,x_2|u^{'},v^{'})  & j = 1,2,\ldots,|\mathcal{X}_1|.|\mathcal{X}_2| - 1 \\
H(Y_2|X_2,V^{'} = v^{'}) & j = |\mathcal{X}_1|.|\mathcal{X}_2| 
\end{array}
\right.
\end{displaymath} 
Similarly, $V^{'}|U^{'} = u^{'}$ suffices to be taken to satisfy $|\mathcal{V}^{''}| \leq |\mathcal{X}_1|.|\mathcal{X}_2|$.
However, the aforementioned Markov chains are still not necessarily satisfied by these RVs. An option to construct new RVs to satisfy the aforementioned Markov chains, which may not be necessary, can be to define,
\begin{IEEEeqnarray}{rCl}
V^{'''} \triangleq (V^{''},X_2)  \qquad \qquad U^{''} \triangleq (U^{'},X_1) \nonumber
\end{IEEEeqnarray}
and the Markov chains are also satisfied and also,
\begin{IEEEeqnarray}{rCl}
I(X_2 ; Y_1 |X_1,U^{''}) & = & I(X_2 ; Y_1 |X_1,U^{'}) = I(X_2 ; Y_1 |X_1,U) \nonumber \\
I(X_1 ; Y_2 |X_2,V^{'''}) & = & I(X_1 ; Y_2 |X_2,V^{''}) = I(X_1 ; Y_2 |X_2,V^{'}) = I(X_1 ; Y_2 |X_2,V) \nonumber
\end{IEEEeqnarray}
Thus we can take
\begin{IEEEeqnarray}{rCl}
|\mathcal{U}| \leq |\mathcal{X}_1|.(|\mathcal{X}_1|.|\mathcal{X}_2| + 1)  \qquad \text{and} \qquad |\mathcal{V}| \leq |\mathcal{X}_1|.|\mathcal{X}_2|^2
\end{IEEEeqnarray}
Now notice that from (\ref{UnifCRdef}) we have,
\begin{IEEEeqnarray}{rCl}
p(S=T) & \geq & k \text{Pr}\{ \Phi = \Psi = l \} \nonumber \\
& = & \text{Pr}\{ \cup_{l \in [1:K]} \{ \Phi = \Psi = l \} \} \geq 1 - \lambda
\end{IEEEeqnarray}
Therefore,
\begin{IEEEeqnarray}{rCl}
P(\Phi \neq \Psi) \leq \lambda
\end{IEEEeqnarray}
Now from Fano's inequality we have,
\begin{IEEEeqnarray}{rCl}
\max\{H(\Psi | \Phi) , H(\Phi | \Psi)\} \leq 1 + \lambda \log(K)   \label{part1}
\end{IEEEeqnarray}
We also have,
\begin{IEEEeqnarray}{rCl} 
H(\Phi,\Psi) & \stackrel{(a)}{\geq} & - \sum_{l=1}^{K}\text{Pr}\{\Phi = \Psi = l\} \log(\text{Pr}\{\Phi = \Psi = l\}) \nonumber \\
& \stackrel{(b)}{\geq} & \sum_{l=1}^{K}\frac{1 - \lambda}{K} \log \left( \frac{K}{1 + \lambda} \right) \nonumber \\
& \geq & (1 - \lambda) \log(K) - 1 \label{part2}
\end{IEEEeqnarray}
where $(a)$ follows from conditioning on the event $A \triangleq \{\Phi = \Psi\}$ and $(b)$ follows from (\ref{UnifCRdef}).
From (\ref{part1}) and (\ref{part2}) we have
\begin{IEEEeqnarray}{rCl}
\min\{ H(\Psi), H(\Phi) \} \geq (1 - 2\lambda) \log(K) - 2 \label{part3}
\end{IEEEeqnarray}
Thus from (\ref{part1}) and (\ref{part3}) we have for every $0 \leq \lambda \leq 1$
\begin{IEEEeqnarray}{rCl}
-3 + (1 - 3\lambda) \log(K) & \leq & H(\Phi , \Psi) \leq H(Y_1^n , Y_2^n) \\
-2 + (1 - 2\lambda) \log(K) & \leq & \min\{H(\Phi) , H(\Psi)\} \leq \min\{H(Y_1^n) , H(Y_2^n)\} \\
-1 + (1 - \lambda) \log(K) & \leq & I(Y_1^n;Y_2^n)
\end{IEEEeqnarray}
and thus we have the following Theorem, 

For instance for every $0 \leq \lambda \leq 1$ we have,
\begin{IEEEeqnarray}{rCl}
-3 + (1 - 3\lambda) \log(K) & \leq & H(Y_1^n , Y_2^n) = n(H(Y_1|X_1,X_2) + H(Y_2|X_1,X_2)) 
\end{IEEEeqnarray}
and thus,
\begin{IEEEeqnarray}{rCl}
\frac{-3}{n} + \frac{(1 - 3\lambda)}{n} \log(K) & \leq & H(Y_1|X_1,X_2) + H(Y_2|X_1,X_2)
\end{IEEEeqnarray}

\begin{theorem}
The CR capacity of the intertwined two-way setting as described in Definition \ref{RDdef} is dominated by,
\begin{IEEEeqnarray}{rCl}
R \leq \min\{ A, B, C, D \} \nonumber
\end{IEEEeqnarray}
where
\begin{IEEEeqnarray}{rCl}
A & \triangleq & H(Y_1|X_1,X_2) + H(Y_2|X_1,X_2) \nonumber \\
B & \triangleq & H(Y_1|X_1,X_2) + I(X_2;Y_1|X_1,U) \nonumber \\
C & \triangleq & H(Y_2|X_1,X_2) + I(X_1;Y_2|X_2,V) \nonumber \\
D & \triangleq & I(X_2;Y_1|X_1,U) + I(X_1;Y_2|X_2,V) \nonumber
\end{IEEEeqnarray}
for some distribution $p(u,v)p(x_1|u)p(x_2|v)$ with $|\mathcal{U}| \leq |\mathcal{X}_1|.(|\mathcal{X}_1|.|\mathcal{X}_2| + 1)$ and $|\mathcal{V}| \leq |\mathcal{X}_1|.|\mathcal{X}_2|^2$.
\end{theorem}
Notice that time-sharing RV $Q$ is used in the achievable rates to convexify (and possibly enlarge) the region. Here, however, we do not need to make the outer bound convex and enlarge it. Hence, there is no point in using a time-sharing RV in the outer bound.
Now let us compare this outer bound with Venkatesan-Anantharam's CR capacity,
\begin{IEEEeqnarray}{rCl}
A & = & H(Y_1|X_1,X_2) + H(Y_2|X_1,X_2) = H(Y_1|X_2) + H(Y_2|X_1) \label{A-up} \\
B & = & H(Y_1|X_1,X_2) + I(X_2;Y_1|X_1,U) \leq H(Y_1|X_2) + I(X_2;Y_1) \label{B-up} \\
C & = & H(Y_2|X_1,X_2) + I(X_1;Y_2|X_2,V) \leq H(Y_2|X_1) + I(X_1;Y_2)  \label{C-up} \\
D & = & I(X_2;Y_1|X_1,U) + I(X_1;Y_2|X_2,V) \leq I(X_2;Y_1) + I(X_1;Y_2) \label{D-up}
\end{IEEEeqnarray}
Therefore, for the case of $p(y_1,y_2|x_1,x_2) = p(y_1|x_2)p(y_2|x_1)$, our outer bound is bounded above as follows,\\
From (\ref{A-up}) and (\ref{B-up}) we have
\begin{IEEEeqnarray}{rCl}
R & \leq & H(Y_1|X_2) + \min \{ H(Y_2|X_1) , I(X_2;Y_1) \} \label{AB-up}
\end{IEEEeqnarray}
and from (\ref{C-up}) and (\ref{D-up}) we have,
\begin{IEEEeqnarray}{rCl}
R & \leq & I(X_1;Y_2) + \min \{ H(Y_2|X_1) , I(X_2;Y_1) \} \label{CD-up}
\end{IEEEeqnarray}
Therefore, from (\ref{AB-up}) and (\ref{CD-up}) we have,
\begin{IEEEeqnarray}{rCl}
R & \leq & \min \{ H(Y_1|X_2) , I(X_1;Y_2) \} + \min \{ H(Y_2|X_1) , I(X_2;Y_1) \}
\end{IEEEeqnarray}
which is the CR capacity of decomposing two-way setting of Anantharam. Therefore, our outer bound is tight in that case and the CR capacity of  RD decomposing setting $p(y_1,y_2|x_1,x_2) = p(y_1|x_1,x_2)p(y_2|x_1,x_2)$ is no more than CR capacity of decomposing setting $p(y_1,y_2|x_1,x_2) = p(y_1|x_2)p(y_2|x_1)$.

\bibliographystyle{IEEEtran}
\bibliography{refs}
\end{document}